\documentstyle[12pt,epsfig,colordvi]{article}

\newcommand{\be}{\begin{eqnarray}}
\newcommand{\ee}{\end{eqnarray}}

\def\anue{{\bar\nu_e}}

\def\gsim{\:\raisebox{-0.5ex}{$\stackrel{\textstyle>}{\sim}$}\:}

\begin{document}

\begin{flushright}
SHEP 02-17     \\
hep-ph/0207260
\end{flushright}

\begin{center}
{\Large \bf Gamma ray bursts as probes of neutrino mass,
quantum gravity and dark energy}
\vspace{.5in}

{\bf Sandhya Choubey and 
S.F. King}
\vskip .5cm

{\it Department of Physics and Astronomy, University of Southampton, \\
Highfield, Southampton S017 1BJ, UK}
\vskip 1in

\end{center}

\begin{abstract}
We calculate the time delays of neutrinos emitted in gamma ray bursts
due to the effects of
neutrino mass and quantum gravity using a time dependent
Hubble constant which can significantly change the naive results presented
hitherto in the literature for large redshifts, and
gives some sensitivity to the details of dark energy.
We show that the effects of neutrino mass, quantum gravity
and dark energy may be
disentangled by using low energy neutrinos to study neutrino mass,
high energy neutrinos to study quantum gravity, and large
redshifts to study dark energy. From low energy neutrinos 
one may obtain
direct limits on neutrino masses of order $10^{-3}$ eV, and
distinguish a neutrino mass hierarchy from an inverted mass hierarchy.
From ultra-high energy neutrinos the sensitivity to the 
scale of quantum gravity can be pushed up to
$E_{QG}\sim 5\times 10^{30}$ GeV. By studying neutrinos from
GRBs at large redshifts a cosmological constant could be distinguished
from quintessence.
\end{abstract}

\newpage

\section{Introduction}

Gamma ray bursts (GRBs) are amongst the most distant, energetic and
enigmatic astrophysical phenomena known. Understanding GRBs is arguably
the most outstanding question in astronomy, and one which may be
answered by a plethora of gamma ray observatories 
such as INTEGRAL, SWIFT and BATSE, and corresponding infrared
and optical telescopes such as REM and LT \cite{GRBreview}.
It is well known that 99\% of the energy of a supernova
is emitted in the form of neutrinos, and therefore it is 
widely expected that GRBs are similarly a copious source of neutrinos
which may be detected in future neutrino telescopes
\cite{Halzen:1996qw, Alvarez-Muniz:2000st, Halzen:1999xc, Gupta:2002zd}.
Within this decade it is
therefore likely that GRBs will become much better understood,
and their exact nature and mechanisms which drive their internal engines
will be revealed. For example it may turn out that a GRB results from
the core collapse of a very massive supernova to a compact rotating black
hole with the energy emitted in beamed relativistic fireball jets
containing copious neutrino fluxes
\cite{Waxman:1997ti, Waxman:1998yy, Bahcall:1999yr, Waxman:1999ai,
Bahcall:2000sa, Halzen:2002pg}
Alternatively the GRB engine could result in the emission of 
beamed earth-sized cannonballs \cite{Rujula:2002xj}.

In this paper we are not concerned with detailed models of GRBs,
but instead regard them as a high intensity, high energy
neutrino beam with a cosmological baseline. 
We shall be interested in
the time delay of the arrival of neutrinos.
We show that
the time delay may be used 
as a probe of three physical effects: (i) neutrino mass, 
(ii) quantum gravity, (iii) dark energy. 
The time delay due to neutrino mass has been noted earlier 
in \cite{Halzen:1996qw} while effects of quantum gravity 
on the time of flight have been considered in 
\cite{Amelino-Camelia:1997gz, 
Biller:1998hg, Schaefer:1998zg, Ellis:1999sd} for 
high energy photons and in 
\cite{Alfaro:1999wd, Ellis:1999sf}
for neutrinos.
However none of the above papers considers the time delays
due to massive neutrinos in the presence of quantum gravity, although
\cite{Alfaro:1999wd} gives the dispersion relation for this case.
Moreover no studies to date have calculated
time delays using a formalism which correctly takes
into account the time dependence of the Hubble constant due to matter
and dark energy. In our study we consider 
the time dependence of the Hubble constant due to matter and dark energy 
and show that they can change the naive results by 
more than 100\% for $z>1$.
For large redshifts the results for time delays due to neutrino
mass and quantum gravity are sensitive to the nature of dark energy.
We show that the three effects may be
disentangled by using low energy neutrinos to study neutrino mass,
high energy neutrinos to study quantum gravity, and using large
redshifts to study dark energy, leading to the results stated in
the abstract.

For the determination of the neutrino mass, 
in principle one could compare the arrival time of the massive neutrinos 
with the arrival times of the photons emitted in the GRB, assuming 
them to be emitted at the same time. However this would be 
plausible only if the GRB were a point source {\it in vacuo}, 
which it is not. In any realistic GRB model the 
photons are trapped inside the 
fireball and are released much later -- the exact amount of time delay 
being highly model dependent. Another strategy might be to 
compare the arrival times of the low energy neutrinos with 
that of the ultra-high energy ones, since (as we will show) 
the ultra-high energy neutrinos suffer negligible time delay due to mass.
But again one would expect the low energy neutrinos to be produced
thermally, as in supernovae, leading to a model dependent time delay. 

However there are alternative strategies which could overcome these
problems. To begin with, if the neutrinos are hierarchical
in mass, $m_1\ll m_2\ll m_3$, then due to their mixing,
neutrinos of the {\em same energy} will arrive at the detector 
in three bunches, corresponding to the three mass eigenstates, 
the arrival time only depending on their mass.
We can then compare the arrival times of the different neutrino
mass eigenstates and put limits on the neutrino mass. 
This represents a ``clean strategy'' and indeed similar 
arguments for constraining neutrino mass from time delays 
using supernova neutrinos have been used in the literature
(see \cite{Beacom:1998yb} and references therein). However we will 
show that with GRB neutrinos the sensitivities possible are 
better by orders of magnitude. 

For constraining models of quantum gravity and dark energy we
can use arrival times of the ultra-high energy neutrinos, which are
produced over many decades of neutrino energy. We can compare the arrival
times of the high energy neutrinos of different energies --
which unlike the low energy thermal neutrinos are expected 
to be produced at almost the same time at the source --
and put limits from such observations. Thus, the results in this paper
do not rely on the comparison of the arrival times
between photons and neutrinos, which would involve the uncertainties
discussed above. However for simplicity, we shall calculate
arrival time delays relative to a hypothetical low energy
photon which is assumed to be emitted at the same time.
This is for convenience only; in a realistic search strategy
what will be important will be the
comparison of time delays between neutrinos of either the same energy, or 
between high energy neutrinos of two different energies as discussed above.

\section{Formalism}
For a neutrino of mass $m$ and energy $E(t)$, the momentum $p(t)$ 
in the presence of the effects of quantum gravity (QG) with an effective energy
scale $E_{QG}$ is given to lowest order as \cite{Alfaro:1999wd}
\be
p^2(t)c^2\approx E^2(t)\left( 1-\xi \frac{E(t)}{E_{QG}} \right) -m^2c^4
\label{dispersion}
\ee
where $\xi=\pm 1$.
It is worth briefly discussing the origin and reliability of the QG
corrections to the dispersion relation in Eq.\ref{dispersion}.
Ellis et al \cite{Ellis:1999sd,Ellis:1999sf} have shown that 
quantum fluctuations in space-time foam lead to a energy dependent 
perturbed gravitational background metric 
\be
g_{ij}=\delta_{ij},~~g_{00}=-1,~~g_{0i}=u_i/c
\label{metric}
\ee
where $|\vec{u}/c|\sim E/E_{QG}$, $E_{QG}$ being the scale at which 
the quantum gravity effects set in. Such a change in the metric  
violates Lorentz invariance at some high scale $E_{QG}$ and changes 
the dispersion relations of the particles. The linear energy dependence 
of such Lorentz invariance violating (LIV) terms in these models 
come mainly from the gravitational recoil effects. An identical 
form for the dispersion relation Eq.\ref{dispersion} is obtained 
by studying massive spin-1/2 fields in the 
framework of loop quantum gravity where the LIV breaking correction term in 
Eq.\ref{dispersion} arises due to the discrete structure of space-time 
at Planck scale \cite{Alfaro:1999wd}. 

In fact the last term in 
Eq.\ref{dispersion} can arise in any theory which violates Lorentz 
invariance. 
The energy dependent form of the corrections may arise from a 
dimension 5 LIV operator of the form 
$c_{\mu \nu
\lambda}\overline{\psi}\gamma_{\mu}D_{\nu}D_{\lambda}\psi$.
The  dimension 4 operator 
$c_{\mu \nu}\overline{\psi}\gamma_{\mu}D_{\nu}\psi$
would lead to an energy independent correction to the dispersion
relation while the dimension 6 operator
$c\overline{\psi}\gamma_{\mu}D_{\nu}D^2\psi$ would lead
to only tiny energy independent corrections.
Hence one may regard the dispersion
relation in Eq.\ref{dispersion} as a consequence of a generic 
type of LIV which may arise in some particular
QG theories. 
From the point of view of measuring neutrino masses, the QG
corrections to the dispersion relation represent a possible 
effect which may threaten to swamp the neutrino mass effect,
and therefore it is important to include the largest imaginable
such effects as we do here.

Assuming Eq.\ref{dispersion}
the time dependent neutrino velocity $v(t)$ is then given to lowest
order in terms of the observed neutrino energy $E_0$ as
\footnote{Strictly $E_0$ is the neutrino energy at the time that the
photons are observed on Earth $E_0=E(t_0)$, but since the neutrinos
arrive a short time later this is to excellent approximation
equal to the observed neutrino energy.}
\be
v(t) = \frac{\partial E(t)}{\partial p(t)}
\approx c \left(1- \frac{\epsilon_0^2}{2}\frac{a^2(t)}{a^2(t_0)}
+\frac{3}{2} \xi\epsilon_0'\frac{a(t_0)}{a(t)}
\right)
\label{velnu2}
\ee
where we have expanded in $\epsilon_0^2\ll 1$ and $\epsilon_0'\ll 1$ 
where 
\be
\epsilon_0=\frac{mc^2}{E_0}, \ \ \epsilon_0'=\frac{E_0}{E_{QG}}.
\ee
The time-dependence of the neutrino velocity $v(t)$ arises from
the expansion of the universe which redshifts the neutrino
de Broglie wavelength $\lambda (t)$, and reduces their momentum
and hence their velocity. The neutrino momentum $p(t)$ is 
related to the cosmic scale factor $a(t)$ through
\be 
\frac{p(t)}{p(t_0)}=\frac{\lambda (t_0)}{\lambda (t)}=
\frac{a(t_0)}{a(t)} 
\ee

Now suppose that low energy photons 
\footnote{As discussed in the Introduction, the time delay is defined
in terms of low energy photons for convenience only. What will
matter in a practical search strategy is the comparison of different
time delays between neutrinos of either the same energy, or 
between high energy neutrinos of two different energies.}
and high energy neutrinos
are emitted from a GRB source at time $t_e$ and the 
low energy photons arrive on Earth 
at time $t_0$ while the neutrinos arrive at time $t_\nu$.
The low energy photons will travel at the speed of light $c$,
while the high energy neutrinos will travel with 
time dependent velocity $v(t)$ which may be smaller than $c$ due
to the finite neutrino mass, and may be smaller or larger than $c$ 
due to the dispersive effects of quantum gravity.
The co-moving distance $\chi$ between the GRB
source and the Earth calculated in terms of the low energy 
photons is given by\footnote{While we have used he low energy photons 
for defining the co-moving distance, Eq.\ref{dist} holds even for 
low energy massless neutrinos which travel at speed $c$.}
\be
\chi &=& \int_{t_e}^{t_0} \frac{c dt}{a(t)}
\label{dist}
\ee
where $a(t)$ is the cosmic scale factor. 
The co-moving distance $\chi$ between the GRB source
and the Earth calculated using the neutrinos which travel freely
with time dependent velocity $v(t)$ is
\be
\chi &=& \int_{t_e}^{t_\nu} \frac{ v(t) dt}{a(t)}
\nonumber \\
&=&\int_{t_e}^{t_0}\frac{v(t) dt}{a(t)}
    +\int_{t_0}^{t_\nu} \frac{ v(t) dt}{a(t)}
\label{distnu}
\ee
The co-moving distance in Eq.\ref{distnu} must be equal to
that calculated using the low energy photons or low energy 
massless neutrinos in Eq.\ref{dist}.

If we now assume that the neutrino velocity and the cosmic scale factor 
do not change much over the time scale $t_0-t_\nu$, then
equating Eq.\ref{distnu} to Eq.\ref{dist} and
using Eq.\ref{velnu2} we find the time delay, to leading order in 
$\epsilon_0^2$ and $\epsilon^\prime_0$,
\be
\Delta t = t_\nu - t_0 
\approx
\frac{\epsilon_0^2}{2}
\int_{t_e}^{t_0}   
\frac{a(t)}{a(t_0)} dt
-\frac{3}{2}\xi \epsilon_0'
\int_{t_e}^{t_0}   
\frac{a^2(t_0)}{a^2(t)} dt
\label{tdel1}
\ee
The time delay in Eq.\ref{tdel1} may be expressed 
in terms of an integral over redshift defined as $1+z'\equiv
a(t_0)/a(t)$, and Hubble constant defined as $H\equiv \dot{a}/{a}$,
\be
\Delta t \approx \frac{\epsilon_0^2}{2} I_2 
-\frac{3}{2}\xi \epsilon_0'I_{-1},
\label{tdel2}
\ee
where $I_n$ are ``redshift moments'' of the inverse Hubble constant,
\be
I_n=\int_0^z \frac{dz^\prime}{(1+z^\prime)^n H(z^\prime)}.
\label{In}
\ee
It is interesting to compare the expression for the
time delay in Eq.\ref{tdel2} to 
the result for the total time taken $T$ for the low energy photon 
or massless neutrino to 
travel from the GRB source to the Earth, 
\footnote{The time $T$ represents how far back in time the GRB took place.
As $z\rightarrow \infty$, $T\rightarrow T_0$ where $T_0$ is the
age of the universe.}
\be
T = t_0-t_e= I_1, 
\label{T}
\ee
It is also interesting to compare to the
result for the proper distance $D$ of the 
GRB source to the Earth at the photon arrival time,
\footnote{
$D$ represents the actual distance of the GRB from the Earth as measured now.
As $z\rightarrow \infty$, $D\rightarrow H_D$ where $H_D$ is the
horizon distance that light could have travelled
since the beginning of the universe, which represents the size of the 
observable universe.}
\be
D = a(t_0) \chi =cI_0.
\label{D}
\ee

The upper limit of the integrals in Eq.\ref{In}
$z$ represents the redshift of the GRB source, where
$1+z=a(t_0)/a(t_e)$, and the Hubble constant 
$H(z)$ for a flat universe is given by
\be
H^2(z)/H_0^2= \Omega_M(1+z)^3+\Omega_{DE}e^{I(z)}
\label{hz}
\ee
where
\be
I(z)=3\int_0^z(1+w(z'))\frac{dz'}{1+z'}
\ee
where the equation of state for the dark energy $w$,
defined as the ratio of its pressure to its density $w\equiv p/\rho$,
in quintessence models has a redshift dependence. For constant $w$ we have
\be
H^2(z)/H_0^2= \Omega_M(1+z)^3+\Omega_{DE}(1+z)^{3(1+w)}
\label{hz1}
\ee
and for a cosmological constant $\Lambda$ with $w=-1$ and
$\Omega_{DE}=\Omega_{\Lambda}$ this reduces to,
\be
H^2(z)/H_0^2= \Omega_M(1+z)^3+\Omega_{\Lambda}.
\label{hz2}
\ee
The present day ratio of matter (M) density to critical
density is $\Omega_M \approx 0.3$, the present day ratio of
dark energy (DE) density to critical density is
$\Omega_{DE} \approx 0.7$ and the Hubble constant 
is $H_0 \approx 70 {\rm kms^{-1}Mpc^{-1}}$. 

We shall consider three different cosmological models
which were recently parametrised in \cite{Corasaniti:2002vg}:
cosmological constant (CONST) with potential $V\sim \Lambda^4$;
supergravity (SUGRA) inspired quintessence field (Q) model 
with potential $V(Q)\sim 1/Q^{11} e^{Q^2/2}$;
inverse (INV) power law quintessence field (Q) model with potential
$V(Q)\sim 1/Q^6$.
The equation of state $w(z)$ for the SUGRA and INV models
does not have a simple analytic form, but it may be
parametrised as discussed in \cite{Corasaniti:2002vg},
and we shall use the parametrisation given there in this paper.
Each of these models has $w(z)$ which varies with $z$ in such a way as
to lead to a ``tracking'' behavior. This implies in particular that the
value of the dark energy density
remains of the same order as the matter density
for large redshifts. By comparison the ratio of the cosmological
constant energy density to the matter density falls
as $1/(1+z)^3$ and rapidly becomes negligible at large redshifts.

\begin{figure}[t]
\centerline{\epsfig{figure=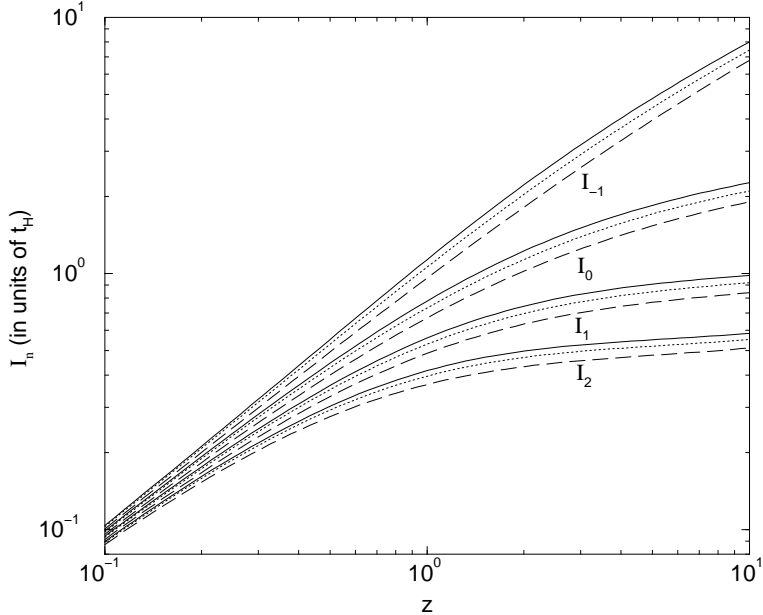,width=4in}}
\caption{The integrals $I_n$ in units of the Hubble time
$t_H\approx 14$ billion years
as a function of the red shift z for three different cosmological models: 
CONST (solid), SUGRA quintessence (dots), INV quintessence (dashes).
$I_0$ is the distance of the GRB in light years, $I_1$ is the date
(years B.C.) that the GRB exploded, $I_2$ is used to calculate the
time delay $\Delta t$ due to neutrino mass,
and $I_{-1}$ is used to calculate the time difference due to 
quantum gravity effects
as discussed in the text.}
\label{in}
\end{figure}

In Figure \ref{in} we calculate $I_n$ in Eq.\ref{In} for $n=-1,0,1,2$
in units of the Hubble time $t_H=1/H_0\approx 14 \ {\rm Gyr}$ as a function 
of redshift $z$ for the three different cosmological models
CONST, SUGRA, INV defined above.
The physically relevant quantities $D$, $T$ and $\Delta t$
are simply related to $I_0$, $I_1$, $I_2$ and $I_{-1}$ as shown in 
Eq.\ref{D},\ref{T},\ref{tdel2}.
Figure \ref{in} shows that as $n$ increases $I_n$ becomes 
increasingly insensitive to redshift (flatter curves) and to 
the particular cosmological model (closer spacing between solid,
dotted and dashed curves) at large redshift.
Both effects can be simply understood as being due to 
the factor of $(1+z^\prime)^n$
in the denominator which tends to suppress the contributions 
from the higher redshift parts of the integration region for larger $n$.
The difference between the curves for $I_0$ and those for $I_{-1}$, $I_2$ 
in Figure \ref{in}
represents the error that would be made if the time delay were
calculated using the naive formula
\be
\Delta t^{\rm naive} \approx 
\left(\frac{\epsilon_0^2}{2}-\frac{3}{2}\xi \epsilon_0'\right) \frac{D}{c}
\approx \left(\frac{\epsilon_0^2}{2}-\frac{3}{2}\xi \epsilon_0'\right) I_0
\label{tdel0}
\ee
which ignores the effect of neutrino redshift as has been done
up till now in the literature
rather than the correct formula for $\Delta t$ in Eq.\ref{tdel2}.
For $z=1$ the error incurred by using the naive formula
can clearly be seen to be of order 100\%, with the error rapidly
growing for larger redshifts. It is also clear from Figure \ref{in}
that there is some sensitivity to the nature of dark energy
for large redshifts.

Finally it is interesting to consider the matter dominated limit of our time
delay result corresponding to $\Omega_M=1$, $\Omega_{DE}=0$.
In this limit the expression for the time delay in Eq.\ref{tdel2} reduces to
\be
\Delta t^{\rm matter} \approx \frac{\epsilon_0^2}{5H_0}
\left[1-\frac{1}{(1+z)^{5/2}}\right]
-\frac{3\xi \epsilon_0'}{H_0}
\left[(1+z)^{1/2} - 1 \right]
\label{limit}
\ee
Time delays due to quantum gravity effects in the matter dominated
limit were also considered in 
\cite{Ellis:1999sd}. However the result quoted there
corresponds to the second term in Eq.\ref{tdel0}
rather than the second term
in Eq.\ref{limit} which correctly takes into account the expansion of the
universe.

\section{Results}

The results in this section
are based on the full formula for $\Delta t$ in Eq.\ref{tdel2}, using
the Hubble constant for a flat universe calculated using Eq.\ref{hz}
for the different dark energy models. 

\begin{figure}
\centerline{\epsfig{figure=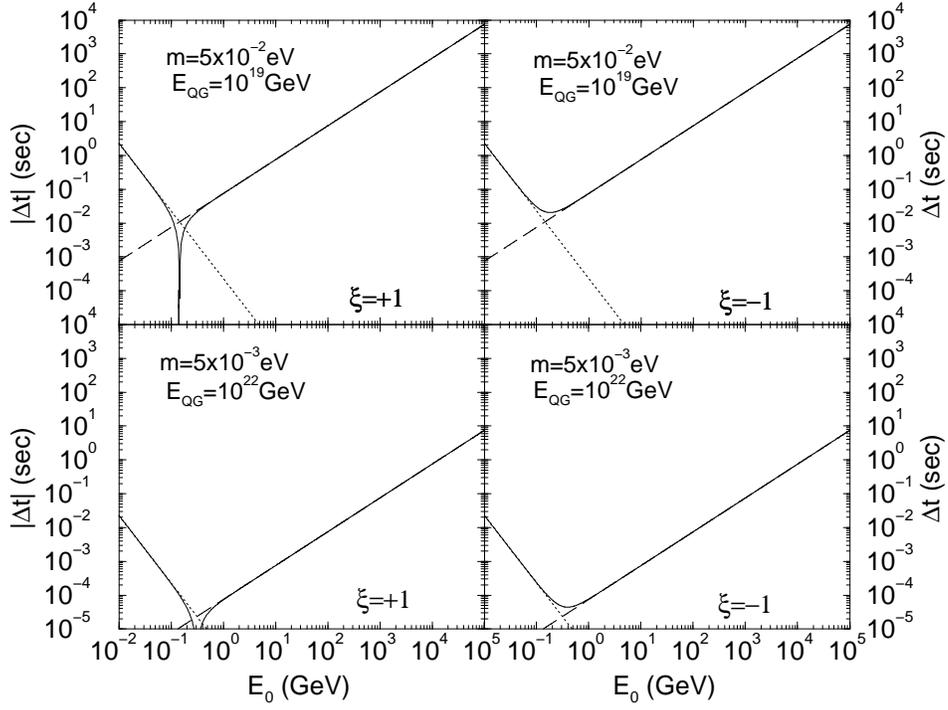,width=5.0in}}
\caption{Time delay $\Delta t$ due to the effect of both 
neutrino mass and quantum gravity against observed neutrino 
energy $E_0$ for $z\sim 1$
for the cosmological constant model. }
\label{delE0}
\end{figure}

Figure \ref{delE0} shows the time delays of the neutrinos
against observed neutrino energy $E_0$ for a fixed GRB redshift
of $z=1$ and assuming the cosmological constant model.
In each panel the downward sloping
dotted lines give the time delay due to 
the effect of neutrino mass which in the upper panels is
chosen to be $m=0.05$ eV corresponding to the ``atmospheric neutrino
mass'' defined as the square root of the atmospheric mass
squared splitting \cite{Fukuda:1998mi}, 
and in the lower panels we take $m=0.005$ eV corresponding 
to the ``solar neutrino mass'' which is the square root
of the large mixing angle solar mass squared splitting 
\cite{Bandyopadhyay:2002xj,Bahcall:2002hv}. To probe neutrino mass 
corresponding to the solar scale the detector has to observe  
neutrinos with $E_0\sim$ few 10 MeV with time sensitivity of
a few milliseconds. The time delays corresponding to the 
atmospheric scale are higher and 
should be easier to detect in the planned neutrino 
telescopes.

The upward sloping dashed lines show the gravitationally induced 
time delay of the neutrinos. 
The upper panels give the time delay when the quantum gravity 
energy scale corresponds to the Planck scale. The lower panels 
are for the case where the quantum gravity corrections become 
important at $10^{22}$ GeV.  
The left panels have
$\xi=+1$ corresponding to a negative time delay due to 
quantum gravity which tends to cancel with the positive time delay
from the effect of neutrino mass. For this case the higher 
energy neutrinos arrive earlier. 
In the figure we have plotted the absolute value of the difference 
between their arrival times.  
The right panels have
$\xi=-1$ corresponding to a positive time delay due to 
quantum gravity which reinforces the positive time delay
from the effect of neutrino mass.
For both the cases we observe that the quantum gavity 
effects become very important 
for neutrinos arriving with energies greater than a few GeV. The 
ultra high energy neutrinos travelling cosmological 
distances can put severe constraints on $E_{QG}$.  
Note that the arguments in \cite{Ellis:1999sf} suggest that 
fermions travel more slowly than $c$ and hence that $\xi=-1$.

The solid lines in the figure show the time delay of the 
neutrinos due to the sum of the dotted and dashed lines
when both the effects are present. For the 
case where $\xi=+1$ the two effects can cancel each other
when from Eq.\ref{tdel2} we have $\Delta t=0$ which occurs at
an energy
\be
E_0'=\left(\frac{m^2 E_{QG}}{3}\frac{I_2}{I_{-1}}\right)^\frac{1}{3}.
\ee
For $\xi=-1$ even though the two effects reinforce each other, 
we see from Eq.\ref{tdel2} and fig. \ref{delE0} that the minima 
in $\Delta t$ comes at exactly the same energy.

Figure \ref{delE0} shows that time delay due to neutrino mass 
is important for the lower energy neutrinos with $E_0<E_0'$,
while the time delay due to quantum gravity effects is important 
for the higher energy neutrinos with $E_0>E_0'$. This enables the
two effects to be disentangled and treated separately.

\subsection{Neutrino Mass Limits from Low Energy Neutrinos}

\begin{figure}
\centerline{\epsfig{figure=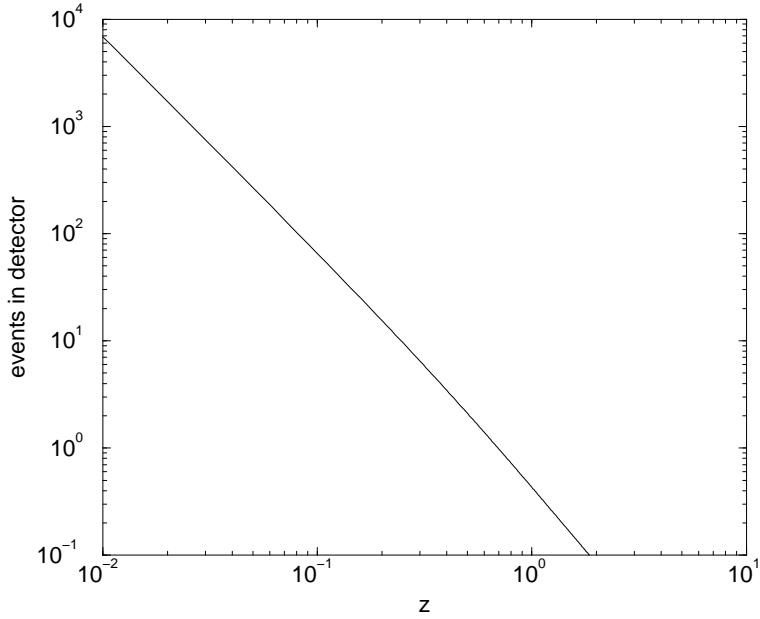,width=4in}}
\caption{Number of events expected from thermal neutrinos 
in a one megaton water 
cerenkov detector as a function of the GRB redshift $z$.}
\label{eventshk}
\end{figure}

In this section we shall focus on
neutrino events with energy $E_0 < E_0'$ in order to 
put limits on $m$. This implies that neutrinos with energies less 
than $\sim 100$ MeV can be effectively used to study their mass. 
Thermal neutrinos with energies between $10-100$ MeV are 
expected to be produced in GRBs \cite{Halzen:1996qw,Halzen:2002pg}. 
Thermal neutrinos, coming from  
GRBs which are energetic enough, which are not very far away and 
which are probably beamed towards the Earth,
should be detectable in the future km$^2$ ice detectors like 
IceCube. These lowest energy 
neutrinos are expected to be detected mainly via
$\bar{\nu}_e+p\rightarrow n+e^+$ where the positrons lead to an
increase in the low PMT noise. This detection method
which forms part of a Supernova Early Warning System may also
be used to detect low energy neutrinos from GRB's 
\cite{Halzen:1996qw,Halzen:2002pg}. However the determination 
of neutrino mass using the time delay techniques requires sensitivity 
to both arrival times as well as energy of the incoming neutrino. 
For a detector like IceCube since the inter-PMT distance is 
large the energy resolution is expected to be poor, particularly for the 
lower energy neutrino events. However energy resolution of the 
proposed megaton Super-Kamiokande type water Cerenkov detectors
such as Hyper-K, UNO and TITAND 
should be good and can be effectively used for putting direct limits 
on neutrino mass. We next calculate the number of events that these 
detectors would observe from a GRB event.

One can estimate the number of thermal neutrinos emitted in a typical 
GRB by the following argument \cite{Halzen:1996qw,Halzen:2002pg}.
Assume that the energy of the GRB in photons is $E_\gamma$ and the radius 
of the GRB fireball is $R$, then the photon temperature $T_\gamma$ is 
given by $T_\gamma = (2E_\gamma/\sigma Vh_\gamma)^{1/4}$, 
where $V=\frac{4}{3}\pi R^3$, $\sigma$ is the Stefan-Boltzmann constant 
and $h_\gamma=2$ is the number of degrees of freedom. Similarly 
the neutrino temperature $T_\nu$ is given by 
$T_\nu = (2E_\nu/\sigma Vh_\nu)^{1/4}$, where 
the number of degrees of freedom for the neutrinos is $h_\nu=2\times 3\times 
\frac{7}{8}$ and $E_\nu$ is the energy emitted 
in neutrinos. We shall assume $E_\nu \approx 100 E_\gamma$
\cite{Halzen:1996qw,Halzen:2002pg}. 
The total number of neutrinos emitted is given by $N_\nu=E_\nu/\langle E_\nu
\rangle$, where $\langle E_\nu\rangle=3.15 T_\nu$ is the average
energy of the neutrinos, assuming a Fermi-Dirac distribution. From this 
we obtain
\be
N_\nu 
= 7 \times 10^{57}
\left(\frac{E_\nu}{10^{54} {\rm ergs}}\right)^{3/4}
\left(\frac{R}{100 {\rm km}}\right)^{3/4}.
\label{total}
\ee
The neutrino flux at Earth is 
\be
\Phi_\nu = \frac{N_\nu}{4\pi D^2}\frac{4\pi}{\Omega_{\rm beam}}
\label{flux_earth}
\ee
where $D$ is the distance of the GRB 
given by Eq.\ref{D} and $\Omega_{\rm beam}$ 
is the beaming solid angle of the GRB 
corresponding to a cone of opening angle 
$\theta_{\rm beam}$, which we assume to be of order $\theta_{\rm beam}
\sim 1^\circ$ \cite{beaming}. 
The total number of events observed at the detector by $\anue$ capture 
on protons  
is given by
\be
N = N_p\int \phi(E)\sigma(E)dE
\label{events}
\ee
where the $\anue$ flux $\phi(E)=\frac{1}{6}\Phi_\nu f_{FD}(E)$, 
$f_{FD}(E)$ being the Fermi-Dirac distribution
function and we assume that the 
total neutrino flux $\Phi_\nu$ is evenly distributed 
in all the six species. The reaction cross-section
$\sigma(E)$ for 
$\bar{\nu}_e+p\rightarrow n+e^+$ is given by 
\be
\sigma(E)=0.94\times 10^{-43} \left(E-Q\right )
\left ( (E-Q)^2 - m_e^2 \right )^{1/2}{\rm cm^2}
\label{cross}
\ee
where $Q=1.293$ MeV is the neutron-proton mass difference and $m_e$ is the 
electron mass. 
$N_p=\frac{2}{18}N_A M_D$ is the number of free protons in water, 
where $N_A$ is the Avogadro's number and $M_D$ is the fiducial mass of the 
detector. The integration in Eq.\ref{events} is done over observed 
neutrino energies.
In Figure \ref{eventshk} we show 
the number of events expected in a one megaton water Cerenkov detector 
from thermal neutrinos produced in a GRB, which is beamed towards 
us with a beaming angle of $1^\circ$ \cite{beaming}. 
The number of expected events are seen to be large for the GRBs 
with low redshift.

\begin{figure}
\centerline{\epsfig{figure=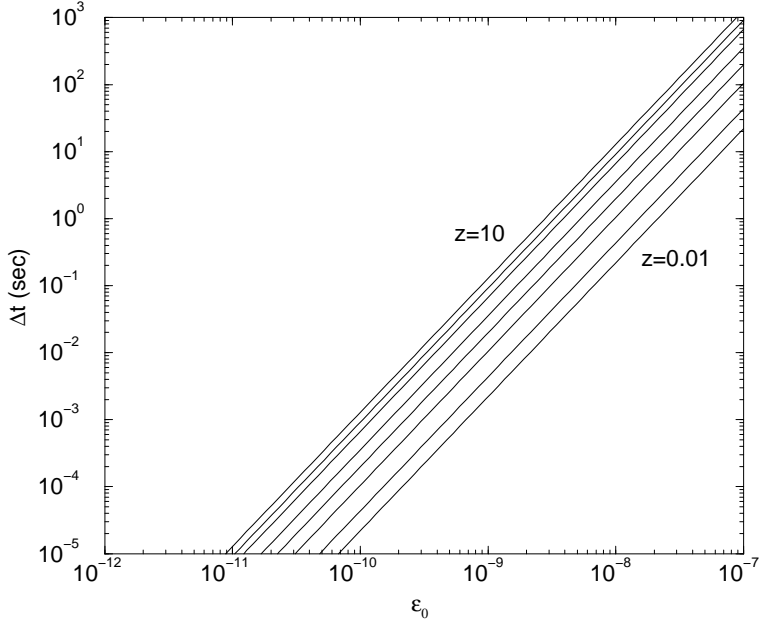,width=4in}}
\caption{ The time delay $\Delta t$ in seconds as a function of 
$\epsilon_0=mc^2/E_0$ for redshift
$z=0.01,0.02,0.05,0.1,0.2,0.5,1,10$, 
assuming the cosmological constant model and taking $\epsilon_0'=0$.}
\label{del1}
\end{figure}

\begin{figure}
\centerline{\epsfig{figure=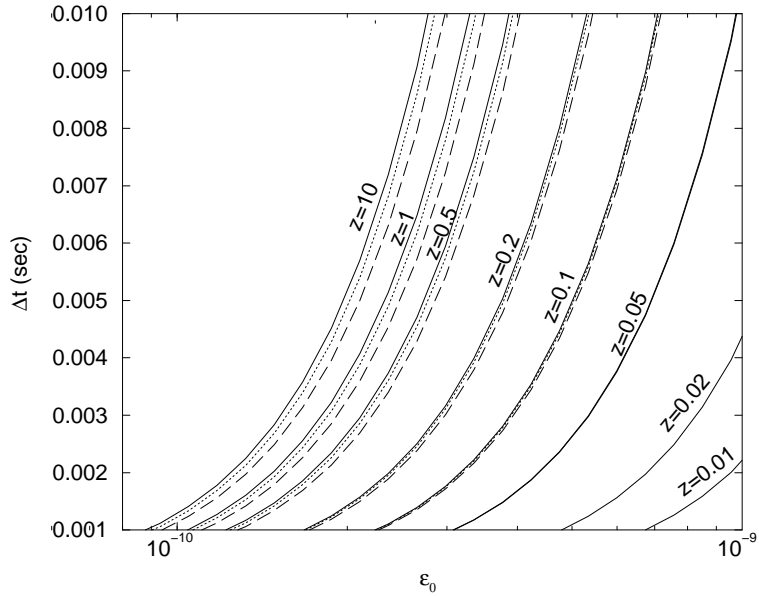,width=4in}}
\caption{Variation in 
time delay $\Delta t$ in the millisecond region
with $\epsilon_0=mc^2/E_0$ for red shift of 
$z=0.01,0.02,0.05,0.1,0.2,0.5,1,10$, 
for the three different cosmological models (taking $\epsilon_0'=0$):
CONST (solid), SUGRA quintessence (dots), INV quintessence (dashes). }
\label{del2}
\end{figure}

In Figure \ref{del1} we show the 
time delay $\Delta t$ in seconds as a function of 
$\epsilon_0=mc^2/E_0$ for redshift
$z=0.01,0.02,0.05,0.1,0.2,0.5,1,10$, 
for the cosmological constant model. 
In this figure we assume that the neutrino energy satisfies
$E_0<E_0'$ so that the effects of quantum gravity can be
neglected and hence we can set $\epsilon_0'=0$.
For a fixed redshift the time delay increases as 
$\epsilon_0^2=m^2c^4/E_0^2$
giving larger time delays for more massive neutrinos
and lower observed energies, as seen in Figure \ref{delE0}.
We are interested in measuring the smallest neutrino
masses, which for a fixed $\epsilon_0$ corresponds to the
smallest observed energies $E_0$ and the smallest time delays.
Since the neutrinos are 
released from the GRB in bursts on the time scale
of milliseconds, the smallest time delays that 
are meaningful will also be milliseconds and from the
figure this corresponds to $\epsilon_0=mc^2/E_0\sim 10^{-10}$. 
For the lowest detectable energy neutrinos
of $E_0\sim 10 {\rm MeV}$, $\epsilon_0\sim 10^{-10}$
corresponds to neutrino
masses of $m\sim 10^{-3} {\rm eV}$. In order to discuss the
limits on neutrino masses in more detail we must focus on
the millisecond region of this plot.

In Figure \ref{del2} we show a blow-up of Figure \ref{del1} 
in the important millisecond region. We have also included
the effect of different cosmological models in this figure,
which can be important for high redshifts.
From Figure \ref{del2} we see that
time delays of $10^{-3} {\rm s}$ correspond to 
$\epsilon_0=7\times 10^{-10}-8\times 10^{-11}$, 
over a range of redshift $z=0.01-10$. For 
the lowest energy neutrinos
of $E_0\sim 10 {\rm MeV}$ this range of $\epsilon_0$ corresponds to
a range of neutrino 
masses of $m= 7\times 10^{-3}-8\times 10^{-4}{\rm eV}$.
Although the smallest neutrino masses are associated with the
highest redshifts
the number of events expected in the detector falls sharply with 
increasing redshift. 
Since $I_{-1}$ is not very sensitive to $z$ beyond $z\sim 1$, 
the fall in the number of events 
with distance is more acute than the rise in the time delay, and hence
GRBs at lower values of redshift would 
be better suited for determination of the neutrino mass.

In addition to being sensitive to the absolute masses of all the three 
neutrino states, GRB neutrinos have the potential to probe the 
mass hierarchy as well.  
Though from solar neutrino 
data we know that the sign of $\Delta m^2_{21}\equiv \Delta m^2_\odot$ 
is positive 
\cite{Bandyopadhyay:2002xj,Bahcall:2002hv}
($\Delta m_{ij}^2=m^2_i-m^2_j$), there is still an ambiguity 
in the sign of $\Delta m^2_{32}\equiv\Delta m^2_{atm}$. 
It would be hard to determine the sign of $\Delta m^2_{32}$ in any of 
the current and planned long baseline oscillation experiments and 
only a neutrino factory would be able to resolve this ambiguity.
The delay in arrival times for the neutrinos can in principle be used  
to determine the neutrino mass hierarchy. 
Neutrinos arrive in three bunches corresponding to the
three mass eigenstates. The neutrinos are detected via the electron
antineutrino reaction, so for each mass eigenstate
we must multiply the detection rate by the probability
$|U_{ei}|^2$ that the mass eigenstate corresponding to the
mass $m_i$ contains an electron antineutrino.
Since $|U_{e3}|^2$ is small $m_3$ will have the smallest
component of electron antineutrino and hence smallest number of events 
at the detector. 
In practice the states with mass $m_1$ and $m_2$ may be most easily detectable
since $|U_{e1}|^2$ and $|U_{e2}|^2$ are not too small. 
The heaviest mass eigenstate will arrive last
and have the largest time delay. 
In hierarchical models
the heaviest mass is $m_3$ while in models with inverted mass 
hierarchy $m_3$ is the lightest. 
So depending on whether fewer events are detected for the 
neutrinos which arrive last (earliest) it would in principle 
be possible to conclude that the mass hierarchy is normal (inverted).


\subsection{Quantum Gravity Limits from High Energy Neutrinos}

\begin{figure}
\centerline{\epsfig{figure=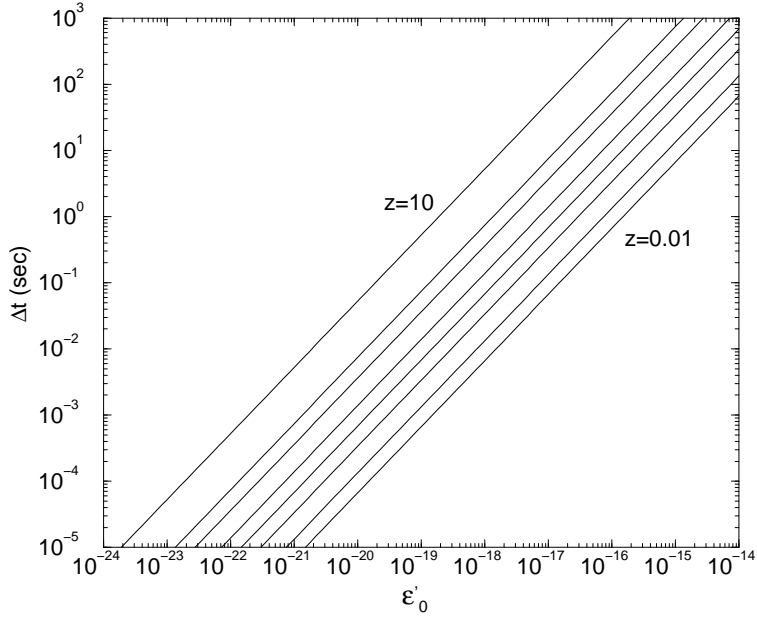,width=4in}}
\caption{ The time delay $\Delta t$ in seconds as a function of 
$\epsilon_0'=E_0/E_{QG}$ for redshift
$z=0.01,0.02,0.05,0.1,0.2,0.5,1,10$, 
assuming the cosmological constant model and taking $\epsilon_0=0$.}
\label{del1p}
\end{figure}

\begin{figure}
\centerline{\epsfig{figure=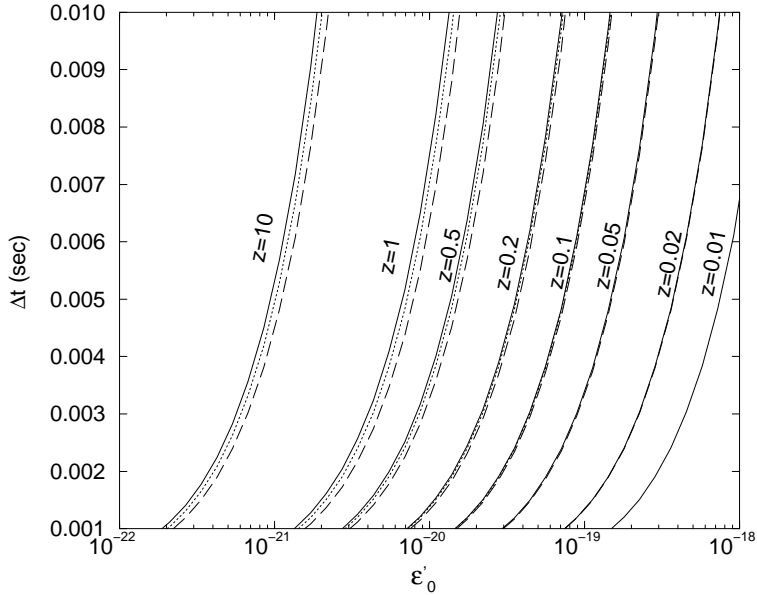,width=4in}}
\caption{Variation in the quantum gravity induced
time delay $\Delta t$ in the millisecond region
with $\epsilon_0'=E_0/E_{QG}$ for red shift of 
$z=0.01,0.02,0.05,0.1,0.2,0.5,1,10$, 
for the three different cosmological models (taking $\epsilon_0=0$):
CONST (solid), SUGRA quintessence (dots), INV quintessence (dashes). }
\label{del2p}
\end{figure}

The dispersion of velocity due to quantum gravity effects
leading to delay in arrival times of the neutrinos is most 
significant at the higher energy end of the GRB neutrino 
spectrum, as evident from Figure \ref{delE0}. In fact for a given 
neutrino mass $m$, all neutrinos with energy 
$E_0>E_0'$ can be used 
to put limits on the scale of quantum gravity $E_{QG}$. Since 
neutrinos with a wide range of energies from $1-10^9$ GeV 
are expected to be produced in GRB fireball 
\cite{Waxman:1997ti,Waxman:1998yy,Bahcall:1999yr,Waxman:1999ai,
Bahcall:2000sa}, 
the detection of their time delays 
is the most powerful method of studying/constraining models which 
predict dispersion relations given by Eq.\ref{dispersion}.

In this section we therefore focus on neutrinos with energy $E_0>E_0'$
in order to set limits on the quantum gravity scale.
Figure \ref{del1p} shows the time delay expected as a function 
of $\epsilon_0^\prime=E_0/E_{QG}$ for redshifts of 
0.01, 0.02, 0.05, 0.1, 0.2, 0.5, 1 and 10, assuming the 
cosmological constant model and also that $E_0>E_0'$
so that we can set $\epsilon_0=0$.
As also seen in Figure \ref{delE0} the time delays 
in Figure \ref{del1p} increase linearly 
with energy and hence with $\epsilon_0^\prime$. The time delay also  
increases significantly with the redshift $z$ as $I_{-1}$ has a 
sharp $z$ dependence. 
Since we want to restrict 
ourselves to minimum time delays of a millisecond, which corresponds 
to the variability time of the GRBs, we show in Figure \ref{del2p} 
the blow up of the region with $\Delta t\sim 10^{-3}$ seconds. 
We also show the effect of the different cosmological models on 
time delays on the same plot.
For $z=0.01-10$ the range 
of $\epsilon_0^\prime$ which would give $\Delta t \leq 10^{-3}$ seconds is 
$\epsilon_0^\prime \leq 1.5\times 10^{-19}-2\times 10^{-22}$. 
For the highest energy neutrinos 
with $E_0\sim 10^{9}$ GeV expected from the GRB, this would translate 
into $E_{QG}\geq6\times 10^{28}-5\times 10^{30}$ GeV. 
This should be compared with the bounds set by the time delays
of the photons observed from GRBs \cite{Amelino-Camelia:1997gz}
where the tightest limit obtained from the highest energy observed 
events is 
$E_{QG} \geq 8.3 \times 10^{16}$ GeV \cite{Schaefer:1998zg}.
In the extreme ultra-relativistic regime 
the effect of neutrino mass is negligible and the dispersion 
of the neutrinos in the quantum space-time foam would be almost the same
as that of photons, the latter having the advantage of being easier 
to detect. However ultra-high energy photons are subject to the 
GZK cut-off and beyond that for better 
bounds on $E_{QG}$ one can use ultra-high energy neutrinos.   
Bounds on $E_{QG}$ from non-observation of dispersion effects
of the quantum space-time foam  
in neutrino oscillations experiments is slightly stronger 
with $E_{QG} \gsim 10^{22}$ GeV \cite{Brustein:2001ik}. This upper 
limit may be improved in the forthcoming long baseline experiments 
using neutrino superbeams. 
But it should be noted that this effect 
will show up in neutrino oscillations only if the 
dispersion due to quantum gravity is 
flavor dependent. However the method using time delays will be sensitive 
even if the effect is flavor independent. Thus 
the observation of time delays of GRB neutrinos 
is the most promising way of phenomenologically testing quantum 
gravity.

\subsection{Dark Energy Limits from High Redshift Events}

\begin{figure}
\centerline{\epsfig{figure=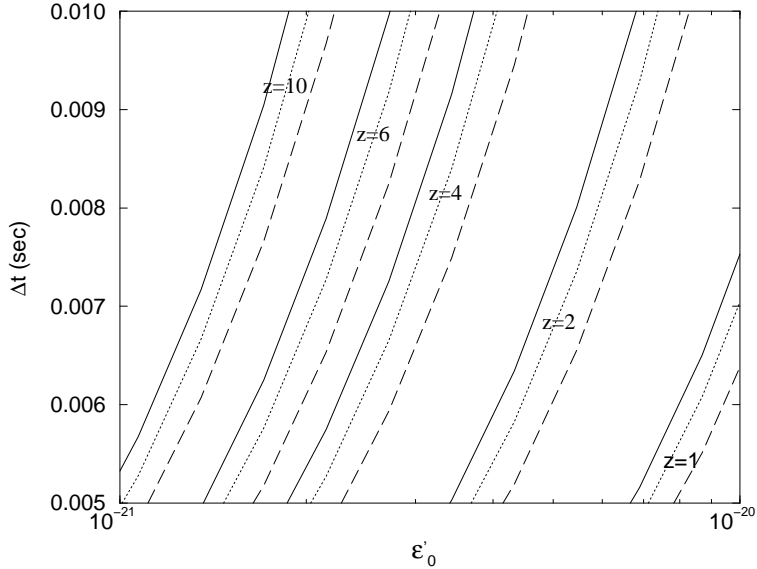,width=4in}}
\caption{Variation in the quantum gravity induced
time delay $\Delta t$ in the millisecond region
with $\epsilon_0'=E_0/E_{QG}$ for high red shifts of 
$z=1,2,4,6,10$ (from right to left)
for the three different cosmological models (taking $\epsilon_0=0$):
CONST (solid), SUGRA quintessence (dots), INV quintessence (dashes).}
\label{del3p}
\end{figure}

Different dark energy models give different values for the Hubble 
constant $H(z)$ at a given redshift. This can have significant impact 
on the time delays. Since the difference in $H(z)$ is maximum at 
the highest redshifts, we have to look at time delays of neutrinos 
coming from the farthest GRB events to probe dark energy.  
Since from Figure \ref{in} we see that 
$I_{-1}$ has a much better sensitivity to dark energy than $I_2$, 
time delays arising from quantum gravity would be a better probe 
for the different cosmological models. Hence it is the ultra high energy 
neutrinos which would be potentially more sensitive. 
In Figure \ref{del3p} we 
show a further blow-up of the time delay plot due to quantum gravity 
effects, with $\Delta t$ between $5\times 10^{-3}-10^{-2}$ seconds and 
$\epsilon_0^\prime$ between $10^{-21}-10^{-20}$. We show the plots for 
$z=$1, 2, 4, 6 and 10. The three different line types 
correspond to the three different models for the dark energy that we 
have considered in this paper. The time delay obtained depends 
on the redshift as well as the cosmological model. 
For $z>3$ the time delays for different redshifts
and different cosmologies begin to overlap. 
This ambiguity can be resolved by using smaller redshift events
to determine $E_{QG}$, since for such events 
the effect of dark energy is reduced. Once $E_{QG}$ is determined then
one can use the high redshift GRB neutrinos to probe dark energy. 
In order to successfully probe the correct cosmological model
it is important that the
redshift of the distant GRBs can be determined accurately enough 
from their afterglow and that the scale of $E_{QG}$ can be ascertained 
well enough by looking at the time delay for neutrinos 
from GRB from lower redshifts. If this can be achieved
then information on $\Delta t$ for neutrinos coming from the very distant 
sources can be used with 
information on $z$ and the scale of $E_{QG}$ to probe the correct 
cosmological model.

\section{Discussion and Conclusion}

In this paper we have calculated the time delays of neutrinos 
emitted in gamma ray bursts due to the effects of
neutrino mass and quantum gravity. Our results
are based on the formula for $\Delta t$ in Eq.\ref{tdel2}, using
the Hubble constant for a flat universe calculated using Eq.\ref{hz}
for different dark energy models. This formalism correctly takes
into account the time dependence of the Hubble constant due to matter
and dark energy, and can change the naive results
in the literature by more than 100\% for $z>1$.
We have shown that the effects of neutrino mass, quantum gravity
and dark energy may be
disentangled by using low energy neutrinos to study neutrino mass,
high energy neutrinos to study quantum gravity, and large
redshifts to study dark energy. From low energy neutrinos 
one may obtain
direct limits on neutrino masses of order $10^{-3}$ eV, and
distinguish a neutrino mass hierarchy from an inverted mass hierarchy.
From ultra-high energy neutrinos the sensitivity to the 
scale of quantum gravity can be pushed up to
$E_{QG}\sim 5\times 10^{30}$ GeV. By studying neutrinos from
GRBs at large redshifts a cosmological constant could be distinguished
from quintessence. 

For convenience we have calculated all time delays with respect
to a hypothetical low energy photon, assumed to be emitted at the
same time from a point source as the neutrino of a given energy.
We emphasise that what is important in practical search strategies
is not these time delays themselves, which will be unmeasurable
due to the uncertainties in the emission characteristics of
low energy photons, but rather the 
comparison of time delays between neutrinos of either the same energy, or 
between high energy neutrinos of two different energies.
As already mentioned, if the neutrinos are hierarchical
in mass then neutrinos of the {\em same energy} will arrive at the detector 
in three bunches, and 
we can then compare the arrival times of the different neutrino
mass eigenstates and put limits on the neutrino mass, 
as in the case of supernova neutrinos 
\cite{Beacom:1998yb} but with sensitivities better by orders of magnitude. 
For constraining models of quantum gravity and dark energy one
can use arrival times of the ultra-high energy neutrinos,
and can compare the arrival
times of the high energy neutrinos of different energies.

\vskip 1in

\begin{center}
{\bf Acknowledgement}
\end{center}
{We particularly acknowledge very helpful 
discussions with Tom Marsh. We also wish to thank 
Juan Garcia-Bellido, Christian Kaiser,
Ian McHardy, Tim Morris and Graham Shore for helpful discussions,
and Francis Halzen for useful communications.}


\end{document}